\documentclass[aps,prc,twocolumn,floats,floatfix,superscriptaddress,nofootinbib]{revtex4}
\usepackage{graphicx,color}
\def\vec#1{\mbox{\boldmath $#1$}}

\begin{document}

\title{Imaginary-time method for radiative capture reaction rate}

\author{K. Yabana}
\affiliation{
Center for Computational Sciences,
University of Tsukuba, Tsukuba 305-8571, Japan }
\affiliation{
Institute of Physics, University of Tsukuba,
Tsukuba 305-8571, Japan}
\affiliation{
Nishina Center for Accelerator-Based Science, The Institute of Physical and Chemical Research (RIKEN), Wako 351-0198, Japan}
\author{Y. Funaki}
\affiliation{
Nishina Center for Accelerator-Based Science, The Institute of Physical and Chemical Research (RIKEN), Wako 351-0198, Japan}
\affiliation{
Institute of Physics, University of Tsukuba,
Tsukuba 305-8571, Japan}

\begin{abstract}
We propose a new computational method for astrophysical reaction 
rate of radiative capture process. 
In the method, an evolution of a wave function is calculated
along the imaginary-time axis which is identified as the inverse
temperature.
It enables direct evaluation of reaction rate as a function 
of temperature without solving any scattering problem.
The method is tested for two-body radiative capture reaction, 
${^{16}{\rm O}}(\alpha,\gamma){^{20}{\rm Ne}}$,
showing that it gives identical results to that calculated by 
the ordinary procedure.
The new method will be suited for calculation of triple-alpha
radiative capture rate for which an explicit construction of 
the scattering solution is difficult.
\end{abstract}

\maketitle

\section{Introduction}

Radiative capture reaction rate far below the Coulomb barrier is an 
essential input for quantitative understanding of stellar evolution 
and nucleosynthesis \cite{Angulo99}. 
However, direct experimental measurements of relevant cross
sections far below the Coulomb barrier often accompany difficulties 
because of their exponentially small cross sections. 
There are also a few three-body processes of significance for which 
experimental measurements are not feasible.
Theoretical evaluation of the radiative capture reaction rate is
thus important.

The radiative capture rate is composed of two distinct contributions,
resonant and nonresonant processes. A potential model is often 
employed for the theoretical evaluation of the nonresonant capture 
rate. For two-body radiative capture processes, it is a routine 
procedure once the model potential is given. 
One first solves the 
radial Schr\"odinger equation  (coupled channel equation if 
necessary) under an appropriate scattering boundary condition 
to obtain scattering cross section. One then calculates the capture 
reaction rate as a function of temperature by integrating the 
cross section over the collision energy with an appropriate 
Boltzmann weight.

Theoretical evaluation of radiative capture reaction rate for
three-body processes is a much more difficult problem.
It is well recognized that the triple-alpha radiative capture reaction to 
form $^{12}$C is a key process to produce heavy elements~\cite{cauldron}. 
At a temperature above 0.1 GK, a resonance state of $0_2^+$
of $^{12}$C which is known as the Hoyle state~\cite{hoyle,fowler} contributes 
dominantly. Below 0.1 GK, on the other hand, nonresonant contribution is 
considered to be significant\cite{Nomoto85,Langanke86,Descouvemont87}. 
Recently, Ogata et.al. 
\cite{Ogata09} conducted a serious evaluation of the rate 
with the CDCC (Continuums-Discretized Coupled-Channels) method, 
a three-body reaction theory which has been successful for 
nuclear direct reactions \cite{cdcc1,cdcc2}.
The radiative capture rate which Ogata et al. reported was 
surprisingly large below $0.1$ GK in comparison with the rate 
which has been employed in standard steller evolution 
calculations \cite{Angulo99}.

Theoretical evaluation of three-body radiative capture rate 
accompanies several difficulties. It is by no means obvious
how to define theoretically the cross section of the triple-alpha 
radiative capture process, because an analytic asymptotic
form of the scattering wave function of three charged-particles
is not known. One also needs to solve the three-body problem
in a huge spatial region for reactions far below the Coulomb 
barrier, since the alpha particles pass through a barrier
for a long distance to penetrate it.

In Ref.~\cite{deDiego11}, de Diego et.al. proposed an alternative
procedure for the calculation of triple-alpha capture rate. 
They consider an inverse process, a photo-absorption of 
$^{12}$C in the excited $2^+$ state, and calculate the transition
probability in the bound state approximation. This procedure
allows one to avoid the difficulty of calculating scattering
solution for three charged-particles. However, a number of 
bound states need to be calculated in their approach.

In this paper, we propose a new computational method for
the radiative capture rate. We will show that the radiative 
capture rate as a function of temperature may be  calculated 
directly by solving an equation which looks like a time-dependent 
Schr\"odinger equation along the imaginary-time axis.
The new method requires neither any solutions of scattering
problem nor any bound state solutions except for a final bound 
state wave function after the capture. 
Since the new method allows us to avoid the difficulties mentioned 
above, we consider it will be useful for the calculation of radiative 
capture rate of triple-alpha process.
In this paper, we demonstrate feasibility of the method by 
applying it to two-body capture reaction, 
${^{16}{\rm O}}(\alpha,\gamma){^{20}{\rm Ne}}$, as an example. 
It will be shown that the new method gives an identical result to 
that calculated by the ordinary method using the two-body 
scattering solution. 

The construction of this paper is as follows. 
In Sec.~\ref{th1}, we present the imaginary-time formalism for
the radiative capture reaction rate. In Sec.~\ref{th2}, we discuss how 
the resonant and nonresonant contributions are included in our 
formalism. In Sec. \ref{tbd}, our method is exemplified by applying
it to the two-body capture reaction, 
$^{16}$O($\alpha$,$\gamma$)$^{20}$Ne.
In Sec.~\ref{tbd1}, we summarize radial equations to be employed
in the practical calculation. In Sec.~\ref{tbd2}, we show results
with ordinary method solving two-body scattering problem. 
We then compare the result of the new method with that of
ordinary method in Sec.~\ref{tbd3}. Sec.~\ref{sm} is devoted 
to summary.

\section{Theory}\label{theory}

\subsection{Imaginary-time method for radiative capture rate}\label{th1}

We consider a radiative capture process of two or three nuclei
confined in a large spatial area of volume $V$.
The transition rate between nuclear states $i$ and  $f$ 
accompanying an emission of a photon of multipolarity 
$\lambda\mu$ is given by \cite{Ring-Schuck},
\begin{equation}
T^{(\lambda\mu)}_{fi}=
\frac{8\pi (\lambda+1)}{\hbar \lambda ((2\lambda+1)!!)^2} 
\left( \frac{E_{\gamma}}{\hbar c} \right)^{2\lambda+1}
\left\vert \left\langle \Psi_f \left\vert
M_{\lambda\mu}
\right\vert \Psi_i \right\rangle \right\vert^2,
\end{equation}
where $M_{\lambda\mu}$ is a transition operator. The energy 
of emitted photon $E_{\gamma}$ is equal to the energy difference 
of two states, $E_{\gamma}=E_i - E_f$. The initial state $\Psi_i$ is a 
scattering state in which $i$ specifies the relative momentum of 
colliding nuclei and other quantum numbers. 
The final state $\Psi_f$ is a bound state wave function after emitting 
the photon.

For two-body collisions, transition rate in a unit spatial area and 
in a unit time under unit number densities of colliding nuclei is given
by $V T^{(\lambda\mu)}_{fi}$ and is equal to $v\sigma_{fi}$, where $v$ is the 
relative velocity and $\sigma_{fi}$ is the cross section. 
For three-body collisions, the reaction rate in a unit spatial 
area and in a unit time is given by $V^2 T^{(\lambda\mu)}_{fi}$.

We denote the inverse temperature as $\beta=1/k_B T$
and express the thermonuclear reaction rate at the inverse
temperature $\beta$ as $r(\beta)$. This is related to the transition 
rate $T^{(\lambda\mu)}_{fi}$ by
\begin{equation}
r(\beta) = 
\frac{\sum_{M_f \mu}\sum_i e^{-\beta E_i} V^{N-1} T^{(\lambda\mu)}_{fi}}
{\sum_i e^{-\beta E_i}}, \label{def_rate}
\end{equation}
where $N=2$ for two-body and $N=3$ for three-body collisions,
respectively. $M_f$ indicates the magnetic quantum number of
final state $f$. The denominator is evaluated to be
\begin{equation}
\sum_i e^{-\beta E_i} \rightarrow
\omega_i \frac{V^{N-1} \mu^{3/2}}
{(2\pi\beta\hbar^2)^{3(N-1)/2}},
\end{equation}
where $\mu$ is the reduced mass, $\mu = m_1m_2/(m_1+m_2)$
for two-body case and $\mu=m_1m_2m_3/(m_1+m_2+m_3)$
for three-body case. $\omega_i$ accounts for the degeneracy
of the initial state.

An essential trick which brings us an imaginary-time evolution
formula for the reaction rate is an employment of the spectral 
representation of the Hamiltonian. Let $f(\hat H)$ be a certain
function of the Hamiltonian operator $\hat H$. We then have
\begin{equation}
f(\hat H) = 
\sum_{n \in bound} f(E_n)\vert \Phi_n \rangle \langle \Phi_n \vert
+ \sum_{i \in scattering} f(E_i) \vert \Phi_i \rangle \langle \Phi_i \vert,
\label{spectral_rep}
\end{equation}
where $E_n$ and $\Phi_n$ are energy eigenvalues and eigenfunctions
of bound states and $E_i$ and $\Phi_i$ are those of scattering states.
For a two-body scattering state, the energy $E_i$ is given by
$E_i = \hbar^2 {\vec k}^2/2\mu$, where $\vec k$ specifies the relative 
wave number of colliding nuclei.

Employing Eq.~(\ref{spectral_rep}) with 
$f(x)=e^{-\beta x}(x-E_f)^{2\lambda+1}$,
one may rewrite Eq.~(\ref{def_rate}) as
\begin{eqnarray}
&&r(\beta) = \frac{1}{\omega_i}
\Big( \frac{2\pi \beta \hbar^2}{\mu} \Big)^{3/2}
\frac{8\pi (\lambda+1)}{\hbar \lambda ((2\lambda+1)!!)^2} \nonumber \\
&& \times \sum_{M_f \mu}
\Big \langle \Psi_f \Big\vert M_{\lambda\mu} e^{-\beta \hat H}
\Big( \frac{\hat H - E_f}{\hbar c} \Big)^{2\lambda+1} 
\hat P M_{\lambda\mu}^{\dagger} \Big\vert \Psi_f \Big\rangle ,
\label{rate_general}
\end{eqnarray}
where $\hat P$ is a projector to remove bound states,
\begin{equation}
\hat P = 1 - \sum_{n \in bound} \vert \Phi_n \rangle \langle \Phi_n \vert .
\end{equation}
Equation (\ref{rate_general}) is the principal result of this paper.
We find that the initial scattering states are removed in this expression.

For a practical calculation of Eq.~(\ref{rate_general}),
we introduce a wave function $\Psi_{\lambda\mu,f}(\beta)$ by
\begin{equation}
\Psi_{\lambda\mu,f}(\beta) = e^{-\beta \hat H}
\left(\frac{\hat H - E_f}{\hbar c} \right)^{2\lambda+1}
\hat P M_{\lambda\mu}^{\dagger} \Psi_f. \label{wf_beta}
\end{equation}
Then the reaction rate is expressed as
\begin{eqnarray}
&&r(\beta) = \frac{1}{\omega_i}
\left( \frac{2\pi \beta \hbar^2}{\mu} \right)^{3/2}
\frac{8\pi (\lambda+1)}{\hbar \lambda ((2\lambda+1)!!)^2} \\
&& \times \sum_{M_f \mu}
\left\langle \Psi_f \left\vert M_{\lambda\mu} 
\right\vert \Psi_{\lambda\mu,f}(\beta) \right\rangle. \label{rc_it}
\end{eqnarray}
The wave function $\Psi_{\lambda\mu,f}(\beta)$ satisfies 
a time-dependent Schr\"odinger equation along the
imaginary-time axis,
\begin{equation}
-\frac{\partial}{\partial \beta} \Psi_{\lambda\mu,f}(\beta)
= \hat H \Psi_{\lambda\mu,f}(\beta),
\label{dpsidbeta}
\end{equation}
with the initial condition,
\begin{equation}
\Psi_{\lambda\mu,f}(0) = \left( \frac{\hat H - E_f}{\hbar c} \right)^{2\lambda+1} 
\hat P M_{\lambda\mu}^{\dagger} \Psi_f .
\label{psi_init}
\end{equation}

In practical calculations, we repeat evolutions with a small 
imaginary-time step $\Delta \beta$ to achieve a finite evolution,
\begin{equation} \Psi_{\lambda\mu,f}(n\Delta \beta)
=\hat P e^{-\Delta \beta \hat H} \hat P \cdots
\hat P e^{-\Delta \beta \hat H} \Psi_{\lambda\mu,f}(0).
\label{dbeta}
\end{equation}
The operation of the evolution operator with a small imaginary-time
step, $e^{-\Delta \beta \hat H}$ may be achieved with the Taylor
expansion method,
\begin{eqnarray}
\Psi_{\lambda\mu,f}(\beta + \Delta \beta)
&=&
\hat P e^{-\Delta \beta \hat H}\Psi_{\lambda\mu,f}(\beta) \nonumber\\
&\simeq&
\hat P \sum_{k=0}^N \frac{(-\Delta \beta \hat H)^k}{k!}
\Psi_{\lambda\mu,f}(\beta).
\label{Taylor}
\end{eqnarray}

In an analytic expression, the projector $\hat P$ is necessary 
only once in Eq.~(\ref{rate_general}), since the Hamiltonian $\hat H$
commutes with the projector $\hat P$. In practical calculations,
however, it is indispensable to apply the projector at each step of
Eq.~(\ref{Taylor}).

\subsection{Resonant and nonresonant contributions}\label{th2}

In the ordinary treatment of radiative capture processes,
contributions of sharp resonances are treated separately 
from the nonresonant contribution. For a two-body collision, 
a contribution of the resonance of energy $E_R$ and width 
$\Gamma$ to the reaction rate is given by \cite{Angulo99}
\begin{equation}
r_R(\beta) =
\left( \frac{2\pi\beta}{\mu} \right)^{3/2}
\hbar^2 \omega_R \frac{\Gamma_i \Gamma_f}{\Gamma}
e^{-\beta E_R},
\label{resonance}
\end{equation}
where $\Gamma_i$ and $\Gamma_f$ are partial widths of 
the resonance to the initial channel through barrier penetration
and to the final state through $\gamma$ emission.
$\omega_R$ is the statistical factor given by
\begin{equation}
\omega_R = \frac{2J_R+1}{(2I_1+1)(2I_2+1)},
\end{equation}
where $J_R$ is the spin of the resonance and $I_{1(2)}$ is
the spin of colliding nucleus 1(2).

Equation (\ref{rate_general}) includes both resonant and
nonresonant contributions since all the final states are
summed up. To confirm that resonant contribution is included
in Eq.~(\ref{rate_general}), we show below that the resonant
contribution $r_R(\beta)$ may be extracted from it.

We assume that the partial decay width for 
gamma emission, $\Gamma_f$, is much smaller than the 
partial width for binary or ternary decay through Coulomb 
barrier, $\Gamma_i$. Indeed, this is the condition that
we may start with the transition rate expression of 
Eq.~(\ref{def_rate}) in perturbation theory. We thus assume that
the partial width decaying into the initial channel, $\Gamma_i$, 
almost exhausts the total width, $\Gamma \simeq \Gamma_i$.
For a sharp resonance, we may express the resonant state
by a normalized wave function $\Phi_R$. 
To calculate the resonant contribution, we replace the
projector $\hat P$ in Eq.~(\ref{rate_general}) with the
projector of the resonant state, 
$\vert \Phi_R \rangle \langle \Phi_R \vert$.
Then we find the contribution of the resonant state may be 
expressed as
\begin{eqnarray}
&&r(\beta; \Phi_R) = \frac{1}{\omega_i}
\left( \frac{2\pi \beta \hbar^2}{\mu} \right)^{3/2}
\frac{8\pi (\lambda+1)}{\hbar \lambda ((2\lambda+1)!!)^2}
e^{-\beta E_R} \nonumber \\
&&\times \left( \frac{E_R-E_f}{\hbar c} \right)^{2\lambda+1}
\sum_{M_f \mu M_R}
\left\vert \left\langle \Psi_f \vert M_{\lambda\mu}
\vert \Phi_R \right\rangle \right\vert^2,
\end{eqnarray}
where $M_R$ specifies a magnetic substate of the resonance.
Noting that the perturbation theory gives an expression
for the radiative decay width of the resonant state $\Phi_R$ as
\begin{eqnarray}
&&\frac{\Gamma_f}{\hbar} = 
\frac{8\pi (\lambda+1)}{\hbar \lambda ((2\lambda+1)!!)^2}
\left( \frac{E_R-E_f}{\hbar c} \right)^{2\lambda+1} \\
&& \times \sum_{M_f \mu}
\left\vert \left\langle \Psi_f \vert M_{\lambda\mu}
\vert \Phi_R \right\rangle \right\vert^2,
\end{eqnarray}
we arrive at the following result.
\begin{equation}
r(\beta; \Phi_R) = \omega_R \left(  \frac{2\pi\beta\hbar^2}{\mu}
\right)^{3/2} e^{-\beta E_R} \frac{\Gamma_f}{\hbar}.
\end{equation}
This is equal to $r_R(\beta)$ if we assume $\Gamma=\Gamma_i$
in Eq.~(\ref{resonance}).

In practical calculations, there are two options when a sharp 
resonance exists. One is to treat the resonant contribution separately
employing Eq.~(\ref{resonance}), removing the resonant 
contribution from the expression of Eq.~(\ref{rate_general}) by adding 
the projector of the resonant state to the projector $\hat P$. 
The other is just to perform the imaginary-time calculation
as it is, so that the resonant contribution is automatically 
included in Eq.~(\ref{resonance}).

\section{Test calculation : $^{16}$O($\alpha,\gamma$)$^{20}$Ne capture rate }\label{tbd}

\subsection{A potential model and radial equations}\label{tbd1}

To confirm that the imaginary-time method explained in the 
previous section works in practice, we apply the method to 
a radiative capture process of two-body collision, 
$^{16}$O($\alpha,\gamma$)$^{20}$Ne.
We assume a simple potential model for the initial $\alpha$-$^{16}$O
scattering state and for the final excited state of $^{20}$Ne. 
This potential model has been adopted in \cite{Mohr05} and 
has been shown to describe the process reasonably.

Numerical calculations will be achieved in the partial wave expansion. 
We first summarize the formula in the partial wave expansion for
$\alpha$-$^{16}$O collision. We introduce a radial wave function
for the relative motion in the ordinary way. For bound states,
we denote
\begin{equation}
\Psi(\vec r) = \frac{u_{nl}(r)}{r}Y_{lm}(\hat {\vec r}),
\end{equation}
where $n$ is the nodal quantum number. We assume a normalization 
relation $\int dr | u_{nl}(r) |^2=1$ as usual.
For scattering states, we denote the radial wave function of
energy $E$ as $u_{El}(r)$ for which we assume the following
normalization in the asymptotic region,
\begin{equation}
u_{El}(r) \rightarrow
\left( \frac{2\mu}{\pi \hbar^2 k} \right)^{\frac{1}{2}}
\sin \left( kr -\frac{l\pi}{2} + \delta_l \right),
\end{equation}
with $E=\hbar^2 k^2/2\mu$. Then, there follows the following 
completeness relation for each $l$ value,
\begin{equation}
\sum_n u_{nl}(r) u_{nl}(r')
+ \int_0^{\infty} dE u_{El}(r) u_{El}(r') = \delta (r-r').
\end{equation}

In the ordinary method, we first calculate the cross section and then
calculate the reaction rate by integrating the cross section
with a Boltzmann weight and a photon phase space factor. 
Denoting the radial wave function of initial state by $u^{(i)}_{E l_i}(r)$
and that of final state by $u^{(f)}_{n_f l_f}(r)$,
the reaction rate is given by
\begin{eqnarray}
&& r(\beta) = \sum_{l_f l_i \lambda}
\frac{2}{\hbar} \left( \frac{2\pi \hbar^2 \beta}{\mu} \right)^{3/2}
\frac{(\lambda+1)(2\lambda+1)}{\lambda((2\lambda+1)!!)^2} \nonumber \\
&& \times 
e^2 \left\{2\left(\frac{16}{20} \right)^{\lambda}
+ 8 \left( -\frac{4}{20}\right)^{\lambda} \right\}^2 \nonumber \\
&& \times (2l_i+1) \langle l_i 0 \lambda_0 \vert l_f 0 \rangle^2
q^{(\lambda)}_{l_f l_i}(\beta),
\end{eqnarray}
where we introduced $q^{(\lambda)}_{l_f l_i}(\beta)$ by
\begin{eqnarray}
&&q^{(\lambda)}_{l_f l_i}(\beta) = 
\int_0^{\infty} dE e^{-\beta E}
\left( \frac{E-E_f}{\hbar c} \right)^{2\lambda+1}  \nonumber \\
&& \times
\left( \int_0^{\infty} dr u^{(f)}_{n_f l_f}(r) r^{\lambda} u^{(i)}_{E l_i}(r) \right)^2.
\label{qdef}
\end{eqnarray}

In the imaginary-time method, we employ a spectral representation
of the Hamiltonian to remove the scattering wave function
$u^{(i)}_{E l_i}(r)$. 
The final expression written in terms of the radial wave function is given by
\begin{equation}
q^{(\lambda)}_{l_f l_i}(\beta)
=
\left\langle u^{(f)}_{n_f l_f} \right\vert
r^{\lambda} e^{-\beta \hat H_{l_i}}
\left( \frac{\hat H_{l_i} - E_f}{\hbar c} \right)^{2\lambda+1}
\hat P_{l_i} r^{\lambda} \left\vert u^{(f)}_{n_f l_f} \right\rangle,
\label{qit}
\end{equation}
where $\hat H_{l_i}$ is the radial Hamiltonian for the partial wave
$l_i$. The $\hat P_{l_i}$ is the radial projector to remove bound
states of the partial wave $l_i$.

Introducing a radial wave function $u^{(\lambda)}_{l_f l_i}(r,\beta)$ by
\begin{equation}
u^{(\lambda)}_{l_f l_i}(r,\beta)
= e^{-\beta \hat H_{l_i}}
\left( \frac{\hat H_{l_i} - E_f}{\hbar c} \right)^{2\lambda+1}
\hat P_{l_i} r^{\lambda} u^{(f)}_{n_f l_f}(r),
\label{uit}
\end{equation}
the function $q^{(\lambda)}_{l_f l_i}(\beta)$ is given by
\begin{equation}
q^{(\lambda)}_{l_f l_i}(\beta)
=
\left\langle u^{(f)}_{n_f l_f} \right\vert
r^{\lambda} \left\vert u^{(\lambda)}_{l_f l_i}(r,\beta) \right\rangle.
\end{equation}

The radiative capture reaction of $^{16}$O($\alpha,\gamma$)$^{20}$Ne 
proceeds dominantly from the initial $\alpha$-$^{16}$O scattering 
state with $l_i=0$ relative angular momentum to the final $2^+$ state 
of $^{20}$Ne at the excitation energy of 1.63 MeV after emission
of $E2$ gamma ray \cite{Mohr05}. 
Since our purpose here is to show the applicability 
of our new method, we concentrate on the calculation of this transition
component. Namely, we consider below the case of $\lambda=2$, 
$l_i=0$, and $l_f=2$.

We assume a simple Woods-Saxon form for the $\alpha$-$^{16}$O
potential with a radius parameter $R_0=2.72$ fm and a 
diffuseness parameter $a=0.85$ fm. 
The depth of the potential is so chosen that the
energies of bound states are reproduced reasonably.
The potential depth of $l=0$ channel is set to $V_0=-150.23$ MeV
to reproduce the ground state energy of $^{20}$Ne from the
$\alpha$-$^{16}$O threshold, $-4.63$ MeV.
The potential depth of $l=2$ channel is set to $V_0=-147.95$ MeV
to reproduce the excitation energy of first $2^+$ state of $^{20}$Ne,
$1.63$ MeV.
There appear many bound states in this potential besides 
the above physical states. They correspond to the 
Pauli-forbidden states of many-body wave function. 
We include all the bound states, both physical and
Pauli-forbidden states, in the projection operator $\hat P_{l_i}$.
The calculations shown below are achieved with a radial grid of 
$\Delta r = 0.1$ fm. The Runge-Kutta method is used to solve
the radial equation from the origin and a simple five-point
finite-difference formula is used in the imaginary-time evolution
for the second-order derivative operator in $\hat H_{l_i}$.

\subsection{Ordinary method}\label{tbd2}

\begin{figure}[htbp]
\includegraphics [scale = 0.70]{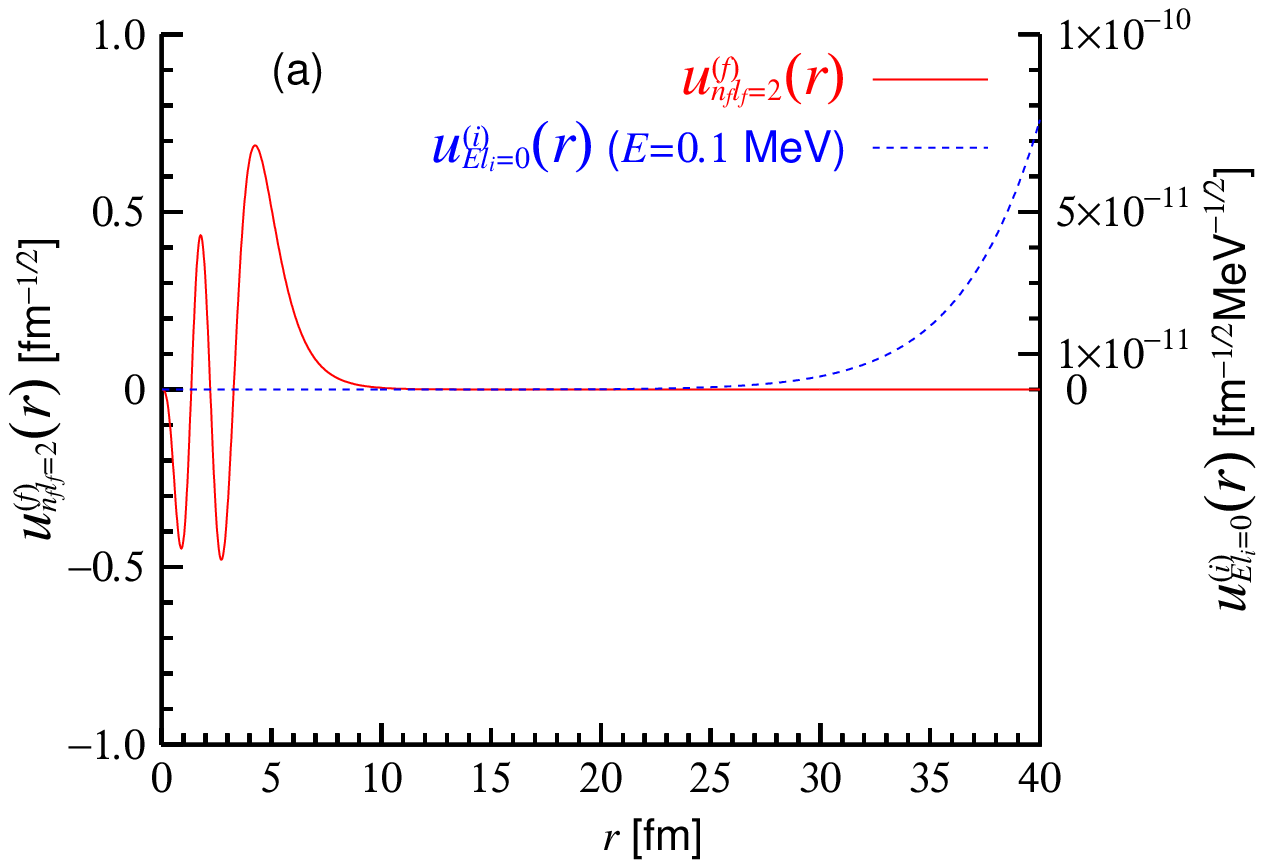}
\includegraphics [scale = 0.70]{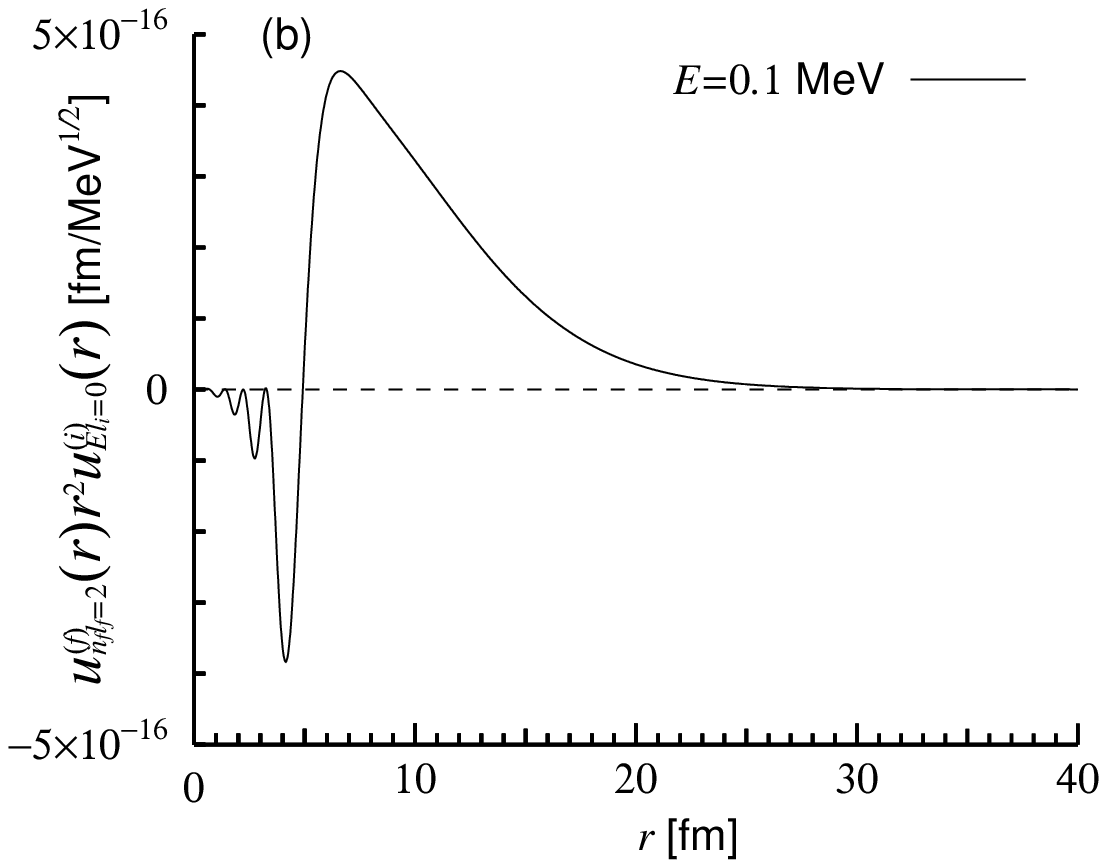}
\caption{\label{radwf} 
(a): The radial wave function of the initial scattering state, $u^{(i)}_{E l_i=0}(r)$,
is shown for the incident relative energy $E=0.1$ MeV by dotted 
curve, and the radial wave function of the final state, $u^{(f)}_{n_f l_f=2}(r)$,
is shown by solid curve.
(b): The overlap function, 
$u^{(f)}_{n_f l_f=2}(r) r^2 u^{(i)}_{E l_i=0}(r)$, 
appearing in the integrand of Eq.~(\ref{qdef}) in the text.
}
\end{figure}

Before showing results with the imaginary-time method, 
we first show calculations in the ordinary approach solving the 
radial Schr\"odinger equation for each incident energy.
Figure \ref{radwf}(a) shows the radial wave functions of initial and 
final states. The initial scattering wave, $u_{E l_i=0}^{(i)}(r)$,
is shown by dashed curve. The incident relative energy is set
to $E=0.1$ MeV, which approximately corresponds to the Gamow
window energy at $T=10^7$K. The final bound-state wave function, 
$u_{n_f l_f=2}^{(f)}(r)$, is shown by solid curve.

Figure \ref{radwf}(b) shows the integrand of the radial matrix
element appearing in Eq.~(\ref{qdef}), 
$u_{n_f l_f=2}^{(f)}(r) r^2 u_{E l_i=0}^{(i)}(r)$. 
As seen from the figure, a dominant contribution comes from
a spatial region where the final wave function $u_{n_f l_f=2}^{(f)}(r)$
decays exponentially. We find the radial integration up to 
30 fm in Eq.~(\ref{qdef}) is required to obtain 
a fully converged result.

\begin{figure}
\includegraphics [scale = 0.75]{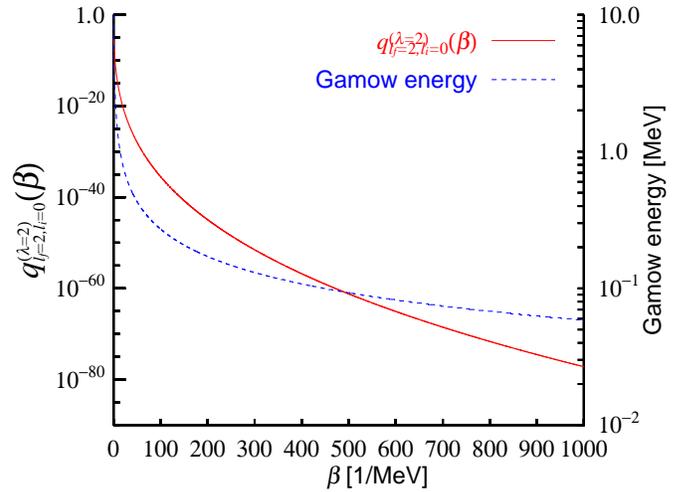}
\caption{\label{rateGamow} 
Solid curve and left scale: radiative capture reaction rate,
$q^{(\lambda=2)}_{l_f=2,l_i=0}(\beta)$,
calculated with the ordinary method of Eq.~(\ref{qdef}).
Dashed curve and right scale: Energy of Gamow window as a 
function of inverse temperature $\beta$.
}
\end{figure}

We show in Fig.~\ref{rateGamow} the reaction rate 
$q^{(\lambda=2)}_{l_f=2,l_i=0}(\beta)$ of Eq.~(\ref{qdef}) as a function of inverse 
temperature $\beta = 1/k_B T$ by solid line (left scale).
We also show the Gamow energy as a function of inverse
temperature by dashed line (right scale). 
The inverse temperature $\beta=10$ MeV$^{-1}$ corresponds 
approximately to $T=1.1 \times 10^9$ K and 
$\beta=1000$ MeV$^{-1}$ to $T=1.1 \times 10^7$ K. 

\subsection{Imaginary-time method}\label{tbd3}

\begin{figure}
\includegraphics [scale = 0.7]{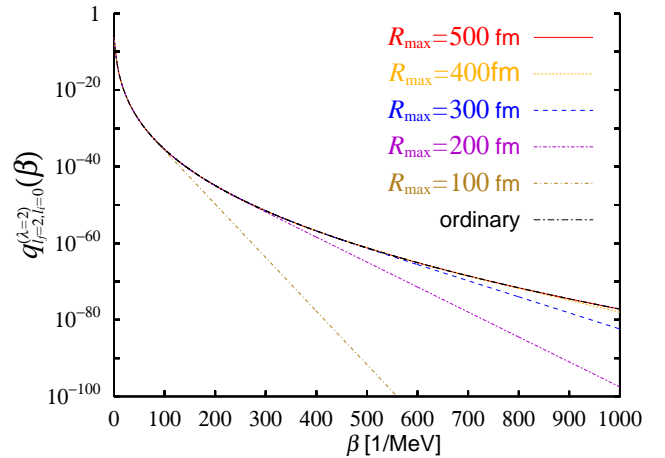}
\caption{\label{rateimag} 
The radiative capture reaction rate $q^{(\lambda=2)}_{l_f=2,l_i=0}(\beta)$
calculated by the imaginary-time method of Eq.~(\ref{qit}) is shown 
for several choices of radial cutoff distance $R_{max}$.
For a comparison, the reaction rate calculated with the ordinary method
of Eq.~(\ref{qdef}) is also shown, which is denoted as ``ordinary''.
}
\end{figure}

We show in Fig.~\ref{rateimag} the reaction rates,
$q^{(\lambda=2)}_{l_f=2,l_i=0}$, calculated with the imaginary-time 
method. We find the calculated reaction rate 
depends on the radial region where the imaginary-time
evolution is calculated. In Fig.~\ref{rateimag}, reaction rates 
calculated with different choices of radial cutoff distance 
$R_{max}$ are compared. The reaction rate in the ordinary
method, which was shown in Fig.~\ref{rateGamow}, is also
shown for comparison.

As is seen in the figure, the reaction rate falls off too rapidly
if the radial cutoff distance is not sufficiently large. The reaction rate 
calculated with the radial cutoff distance $R_{max}=500$ fm
coincides almost completely with the reaction rate calculated 
in the ordinary method for a whole temperature region shown in
the figure. 
We thus conclude that, to obtain accurate reaction 
rate at low temperature with the imaginary-time method, 
one needs to calculate the imaginary-time evolution of the
wave function in a sufficiently large radial space,  up to $500$ fm 
for $T \sim 10^7$ K.

\begin{figure}
\includegraphics [scale = 0.65]{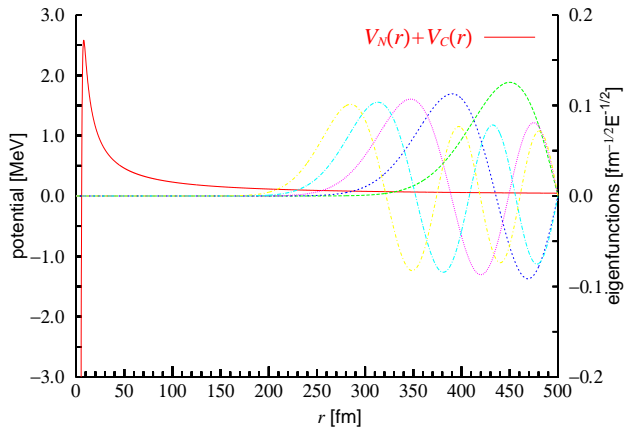}
\caption{\label{potwf} 
The $\alpha$-$^{16}$O potential is denoted by a red solid 
curve (left scale). Five lowest positive-energy eigenfunctions 
of the potential are shown by dotted curves. Calculations
are achieved in the radial region up to $500$ fm.}
\end{figure}

Figure \ref{rateimag} indicates that the reaction rates 
by the imaginary-time method decrease exponentially 
at large $\beta$ when the radial cutoff distance is not 
sufficiently large. We express the asymptotic behavior as 
$q^{(\lambda=2)}_{l_f=2,l_i=0}(\beta) \simeq e^{-\beta \epsilon}$, 
where the slope constant $\epsilon$ depends on the radial 
cutoff distance, $R_{max}$. The $\epsilon$ increases as the
radial cutoff distance decreases. 
In the imaginary-time calculation, the wave function
$u^{(\lambda)}_{l_f l_i}(r,\beta)$ is dominated by the eigenfunction of the
lowest eigenvalue when $\beta$ is sufficiently large.
Since all the bound states are removed in the imaginary-time
evolution by the projection operator, the slope parameter 
$\epsilon$ should coincide with the lowest positive 
energy-eigenvalue of the Hamiltonian.
Since the Coulomb potential decreases monotonically
as a function of radial coordinate,
the eigenfunction of the lowest positive eigenvalue should 
localize in the region close to the radial cutoff distance,
if there is not a sharp resonant state below that energy.

In Fig.~\ref{potwf}, we show the $\alpha$-$^{16}$O potential
and the eigenfunctions belonging to several positive 
low-energy eigenvalues. Calculation is achieved in the 
radial region up to $R_{max}=500$ fm. 
The $\alpha$-$^{16}$O potential is composed of nuclear $(V_N)$
and Coulomb $(V_C)$ potentials, and the lowest positive-energy
eigenvalue is close to the minimum of the Coulomb potential 
energy at the radial cutoff distance, 
$E_{min} \sim Z_1 Z_2 e^2/R_{max}$.
For $R_{max}=500$ fm, the energy is $E_{min} \sim 0.046$ MeV.
For $R_{max}=100$ and $200$ fm, $E_{min} \sim 0.23$ and
$0.115$ MeV, respectively. These values explain the slope of 
the reaction rate at large $\beta$ seen in Fig.~\ref{rateimag}.

It is evident that the imaginary-time evolution in the radial 
region inside a certain radial cutoff distance $R_{max}$ takes 
only into account the tunneling process of energy higher 
than $e^2 Z_1 Z_2/R_{max}$.
As the temperature becomes lower, one needs to calculate 
the reaction rates in a wider radial region. 
We may estimate the radial cutoff distance $R_{max}$ to obtain 
a reliable reaction rate for a given temperature $\beta$
considering the energy of the Gamow window.
Employing a standard formula for the peak energy of the Gamow 
window as a function of temperature and equating the energy 
with the Coulomb potential energy at the radial cutoff distance,
we obtain
\begin{equation}
R_{max} \sim \left( 
\frac{2\hbar^2 Z_1 Z_2 e^2 \beta^2}{\mu \pi^2} \right)^{\frac{1}{3}}.
\end{equation}
This  gives $R_{max} = 85$ fm for $\beta=10^2$ MeV$^{-1}$
and $R_{max} = 394$ fm for $\beta=10^3$ MeV$^{-1}$.
This estimation coincides with the observation in Fig.~\ref{rateimag} 
that the calculation up to $R_{max}=100$ fm describes reaction rate
for $\beta < 100$ MeV$^{-1}$ and the calculation up to
$R_{max}=500$ fm for $\beta < 1000$ fm.

\begin{figure}
\includegraphics [scale = 0.65]{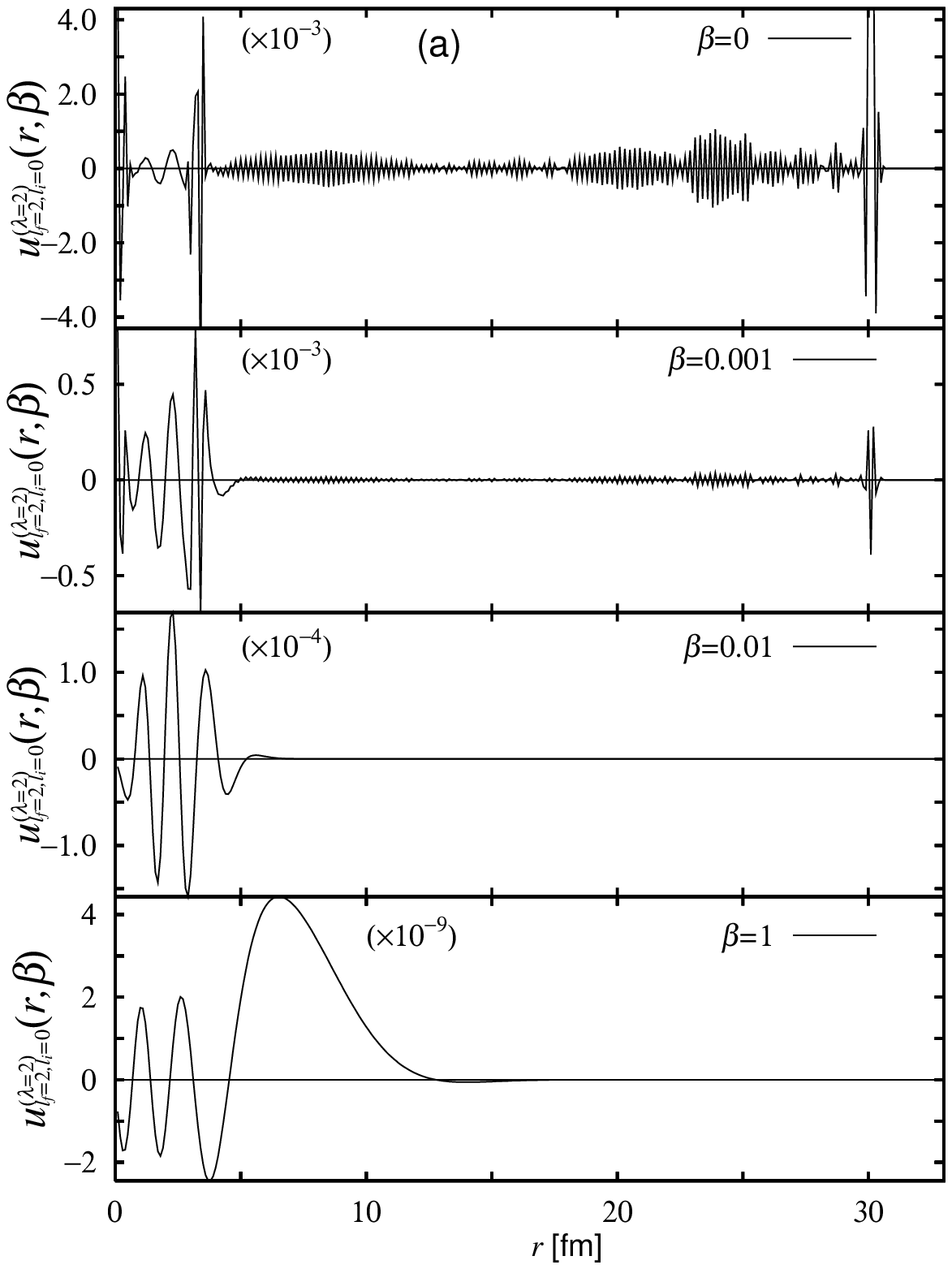}
\includegraphics [scale = 0.65]{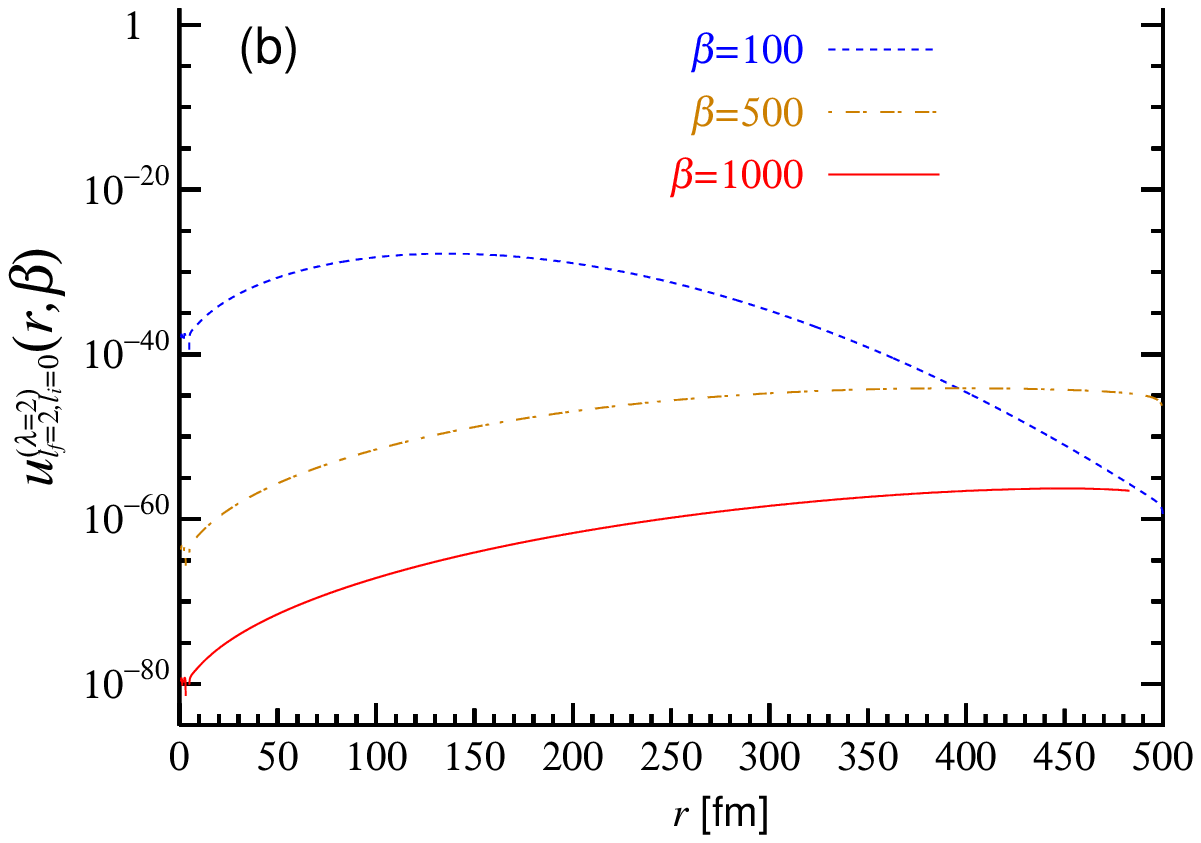}
\caption{\label{imagwf} 
Imaginary-time evolution of the wave function as a function 
of radial coordinate $r$, $u^{(\lambda=2)}_{l_f=2,l_i=0}(r,\beta)$ 
defined by Eq.~(\ref{uit}), (a) at 
$\beta=0,0.001,0.01,$ and $1$ MeV$^{-1}$ in the linear scale 
and (b) at $\beta=100,500,1000$ MeV$^{-1}$ in the logarithmic scale.
}
\end{figure}

For a deeper understanding of the imaginary-time method,
we show in Fig.~\ref{imagwf}(a)(b) the evolution of the radial 
wave function $u^{(\lambda=2)}_{l_f=2,l_i=0}(r,\beta)$ in the
imaginary-time for several values of $\beta$. The calculation 
is achieved with $R_{max}=500$ fm. 
In the top four panels, the wave functions of 
$\beta \leq 1$ MeV$^{-1}$ are shown in the linear scale. 
In the bottom panel, the absolute value of the wave functions are shown for
large $\beta$ values in logarithmic scale.

To start the calculation, we prepare the radial wave function of the 
final state $u_{n_f l_f=2}^{(f)}(r)$ inside
a region, $R_{max}^{(f)}=30$ fm. 
In the top panel of Fig.~\ref{imagwf}(a), the wave 
function at $\beta=0$ is shown. We find a number of 
spikes in the initial wave function 
$u^{(\lambda=2)}_{l_f=2,l_i=0}(r,\beta=0)$. 
In particular, an intense spike is seen at around
$r\simeq R_{max}^{(f)}=30$ fm. 
These sharp structures originate from the operation
of $(\hat H_{l_i}-E_f)^5$ in preparing the wave function at
$\beta=0$. It works to emphasize components with high 
energy eigenvalues of the radial Hamiltonian. 

At first sight, the existence of such sharp structures looks 
unfavorable and problematic. However, these spikes
disappear immediately after we start the imaginary-time
evolution and they will not affect the reaction rate at low
temperature. Even at $\beta=0.001$ MeV$^{-1}$, 
these spikes are substantially reduced. They 
disappear almost completely at $\beta=0.01$.  

As the inverse temperature $\beta$ increases, the wave function
starts to shift outwards. At $\beta=1$ MeV$^{-1}$, the amplitude 
of the wave function shows a peak at around $6$ fm. 
As seen in Fig.~\ref{imagwf}(b), the dominant component of the
wave function gradually shifts towards a region of large radial
distance. Eventually, at $\beta>500$ MeV$^{-1}$,
the radial wave function is dominated in the region of large radial
distance. At $\beta=1000$ MeV$^{-1}$, the wave function of 
small radial region ($r<30$ fm), which contributes to the radiative 
capture rate, is much smaller than that
in the asymptotic region by about $10^{20}$ order of magnitude. 
Thus, the imaginary-time calculation should be achieved with high
accuracy to describe $10^{20}$ difference of magnitude of the 
wave function in different radial regions.

To confirm that the result does not depend on the radial region
in which we prepare the final wave function, we compare
reaction rates employing final wave functions $u^{(f)}_{n_f,l_f}(r)$
prepared in the radial region with different cutoff radius,
$R^{(f)}_{max}$.

\begin{figure}[htbp]
\includegraphics [scale = 0.75]{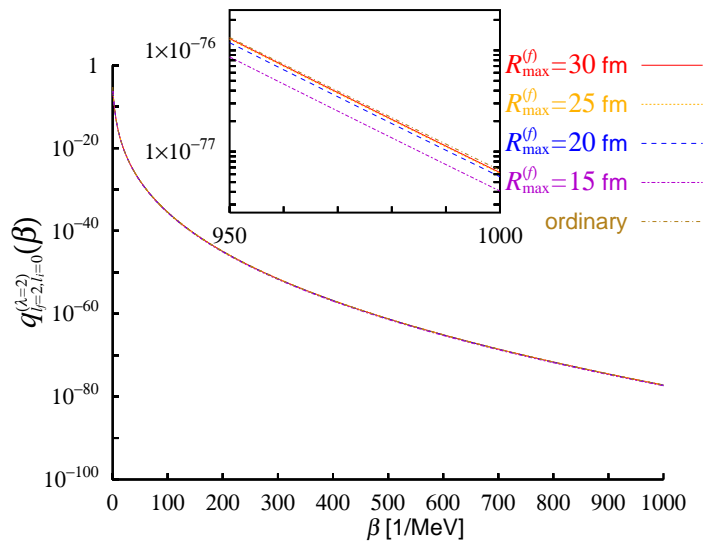}
\caption{\label{radcutoff}
The dependence of the reaction rate $q^{(\lambda=2)}_{l_f=2,l_i=0}(\beta)$ on the radial 
cutoff distance, $R^{(f)}_{max}$, in preparing the final-state wave 
function $u^{(f)}_{n_f l_f=2}(r)$.}
\end{figure}

In Fig.~\ref{radcutoff}, we compare reaction rates calculated by the 
imaginary-time method employing  final wave functions of different 
radial cutoff distance, $R_{max}^{(f)}$.
We find the calculated reaction rate is quite insensitive to the radial 
cutoff distance. We thus confirm that a number of sharp structures
seen in the top panel of Fig.~\ref{imagwf}(a), especially prominent 
at around the radial cutoff distance, $R_{max}^{(f)}$, 
do not have any influence on the reaction rate calculation.
As seen in the inset, the reaction rate is
convergent if we choose $R^{(f)}_{max}\geq 25$ fm, which is
consistent with our observation in Sec.~\ref{tbd1}.

\begin{figure}[htbp]
\includegraphics [scale = 0.75]{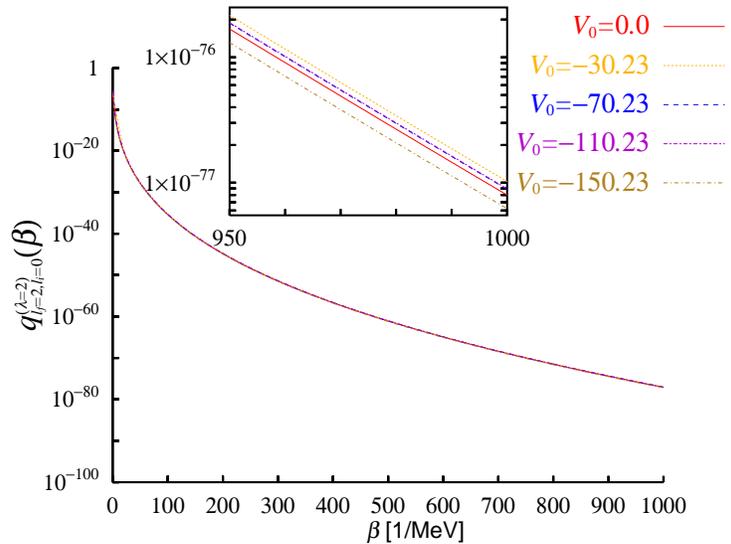}
\caption{\label{potdep}
Reaction rates with different nuclear potential in the
initial scattering channel are shown. The depths of the
Woods-Saxon potential, $V_0$ is varied.}
\end{figure}

We finally present a note on the dependence of the reaction rate 
on the choice of the nuclear potential in the imaginary-time evolution. 
In Fig.~\ref{potdep}, we compare the reaction rates changing the
depths of the nuclear potential, $V_0$ of the Woods-Saxon
potential in the initial scattering state with $l_i=0$. All the other
parameters are set to be the same.
As seen from the figure, the reaction rate is quite
insensitive to the choice of the parameter $V_0$.
Even without the nuclear potential, i.e. with $V_0=0$, the reaction
rate is given almost correctly. As the inset shows, the difference is
within a factor of $1.5$ in 
$950\ {\rm MeV}^{-1}< \beta < 1000\ {\rm MeV}^{-1}$. 
We thus conclude that the nuclear potential in the initial channel,
which will be used in the imaginary-time evolution, has
very small effect on the reaction rate. Of course, this conclusion
applies only to the nonresonant contribution. The resonance energy
and width are sensitive to the nuclear potential, and so is the
resonant contribution to the reaction rate.

\section{Summary}\label{sm}

In this paper, we proposed a new computational method for
radiative capture reaction rate. Employing a spectral representation
of the Hamiltonian, we have shown that the reaction rate as a function
of temperature may be calculated without solving any scattering
problem. Starting with an initial wave function which includes the
final bound-state wave function after the emission of photon, the 
reaction rate as a function of inverse temperature, $\beta (=1/k_B T)$, 
can be obtained directly by solving a time-dependent Schr\"odinger 
equation in the imaginary-time axis.

To show feasibility of the method, 
we show application of the method to 
${^{16}{\rm O}}(\alpha,\gamma){^{20}{\rm Ne}}$ 
reaction in a simple potential model.
We have confirmed that the new method gives an accurate reaction rate 
if we solve the imaginary-time evolution equation 
in a sufficiently large spatial area. 
Since the new method does not require any solution of
scattering equation, it will be a promising approach for 
the reaction rate of triple-alpha radiative capture process.
The application to that process is now in progress.

\end{document}